\documentclass[a4paper]{jpconf}
\usepackage{graphicx}

\usepackage[latin1]{inputenc}

\usepackage{graphicx}
\usepackage{color,pst-plot}

\usepackage{amsmath}
\usepackage {amsfonts,amssymb,amstext}
\usepackage{psfrag}

\newcommand{\beq}[1]{
\begin{equation}\label{#1}}
\newcommand{\eeq}{\end{equation}}
\newcommand{\bea}[1]{
\marginpar{\small\textsf{#1}}
\begin{eqnarray}\label{#1}}
\newcommand{\eea}{\end{eqnarray}}
\def\pv{\vec{p}_t}
\def\dv{\vec{\Delta}_t}

\def\ar{\alpha_\rho}

\def\mr{m_\rho}

\def\bea{\begin{eqnarray}}
\def\beqa{\begin{eqnarray}}
\def\eea{\end{eqnarray}}
\def\eqa{\end{eqnarray}}
\def\beas{\begin{eqnarray*}}
\def\eeas{\end{eqnarray*}}
\def\beqas{\begin{eqnarray*}}
\def\eqas{\end{eqnarray*}}
\def\beq{\begin{equation}}
\def\eeq{\end{equation}}
\def\beqd{\begin{displaymath}}
\def\eeqd{\end{displaymath}}
\def\eqd{\end{displaymath}}

\def\beeq{\begin{eqnarray}} \def\eeeq{\end{eqnarray}}

\newcommand{\eq}{\end{equation}}

\newcommand{\be}{\begin{equation}}
\newcommand{\ee}{\end{equation}}

\DeclareMathAlphabet{\eusm}{U}{}{}{}
\SetMathAlphabet\eusm{normal}{U}{eus}{m}{n}
\SetMathAlphabet\eusm{bold}{U}{eus}{b}{n}
\DeclareMathAlphabet{\mathpzc}{OT1}{pzc}{m}{it}

\input epsf

\def\slashchar#1{\setbox0=\hbox{$#1$}
  \dimen0=\wd0
  \setbox1=\hbox{/} \dimen1=\wd1
  \ifdim\dimen0>\dimen1
     \rlap{\hbox to \dimen0{\hfil/\hfil}}
     #1
  \else
     \rlap{\hbox to \dimen1{\hfil$#1$\hfil}}
     /
  \fi}

\begin{document}
\title{On $\gamma N \to \gamma \rho N'$ at large $\gamma \rho$ invariant mass }

\author{R. Boussarie$^1$, B. Pire $^{2}$, L. Szymanowski$^{1,2,3}$ and S. Wallon$^{1,4}$}

\address{$^1$ Laboratoire de Physique Th\'{e}orique, CNRS, Univ. Paris Sud, Universit\'{e} Paris-Saclay, \hspace*{0.25cm}91405 Orsay, France}
\address{$^2$ Centre de Physique Th\'eorique, Ecole Polytechnique, CNRS, Universit\'e Paris-Saclay, \hspace*{0.25cm}F91128 Palaiseau, France}
\address{$^3$ National Centre for Nuclear Research, Warsaw, Poland}
\address{$^4$ UPMC Universit\'{e} Paris 6, Facult\'{e} de physique, 4 place Jussieu, 75252 Paris Cedex 05,\\ \hspace*{0.25cm}France}

\ead{renaud.boussarie@th.u-psud.fr, bernard.pire@polytechnique.edu, lech.szymanowski@ncbj.gov.pl and samuel.wallon@th.u-psud.fr}

\begin{abstract}
Photoproduction of a pair of particles with large invariant mass is a natural extension of collinear QCD factorization theorems which have been much studied for deeply virtual Compton scattering and deeply virtual meson production. We consider the case where the wide angle Compton scattering subprocess $\gamma (q\bar q) \to \gamma \rho $ factorizes from generalized parton distribution.  At dominant twist, separating the  transverse  (respectively longitudinal) polarization of the $\rho$ meson allows one to get access to  chiral-odd (respectively  chiral-even) GPDs. This opens a new way to the extraction of these elusive transversity GPDs.
\end{abstract}

\section{Introduction}
The last two decades have witnessed great progresses  on the QCD description of hard exclusive processes, in terms of generalized parton distributions (GPDs) \cite{GPD} describing the 3-dimensional content of hadrons \cite{impact}. Much hope exists on a meaningful extraction of dominant (i.e. chiral-even) GPDs from JLab12 near future experiments. To increase our confidence in this extraction, one however needs to probe various processes, thus verifying the universality of the GPDs.   On the other hand, access to the chiral-odd transversity GPDs~\cite{defDiehl}, noted  $H_T$, $E_T$, $\tilde{H}_T$, $\tilde{E}_T$, which decouple from deeply virtual Compton scattering and deeply virtual meson production at leading order, has  turned out to be even more challenging~\cite{DGP} than the usual transversity distributions: one photon or one meson electroproduction leading twist amplitudes are insensitive to transversity GPDs.  Quark mass effects \cite{PS2015} or production of a meson described by a twist 3 
distribution amplitude \cite{liuti} are two ways to evade this difficulty. The alternate strategy followed  in Ref.~\cite{IPST,PLB}, was to study the leading twist contribution to processes where more mesons (denoted A and B) are present in the final state. The hard scale which allows to probe the short distance structure of the nucleon
is  the invariant squared mass of the meson pair $s=M_{A,B}^2\, \sim |t'|$ in the fixed angle regime.
A similar strategy has also been advocated in Ref.~\cite{kumano} for chiral-even GPDs. 

We study  the process:
\begin{equation}
\gamma ^{(*)}(q)+ N (p_1) \rightarrow \gamma(k) + \rho (p_\rho, \epsilon_\rho) + N' (p_2)\,,
\label{process}
\end{equation}
where $\epsilon_\rho$ is the polarization vector of the $\rho$ meson.
This process is  sensitive to both chiral-even and chiral-odd GPDs due to the chiral-even (resp. chiral-odd) character of the leading twist distribution amplitude (DA) of $\rho_L$ (resp. $\rho_T$). To study this process, we closely follow the method described in Ref.~\cite{PLB}.

\begin{figure}[h]

\psfrag{TH}{$\Large T_H$}
\psfrag{Pi}{$\pi$}
\psfrag{P1}{$\,\phi$}
\psfrag{P2}{$\,\phi$}
\psfrag{Phi}{$\,\phi$}
\psfrag{Rho}{$\rho$}
\psfrag{tp}{$t'$}
\psfrag{s}{$s$}
\psfrag{x1}{$\!\!\!\!\!\!x+\xi$}
\psfrag{x2}{$\!x-\xi$}
\psfrag{RhoT}{$\rho_T$}
\psfrag{t}{$t$}
\psfrag{N}{$N$}
\psfrag{Np}{$N'$}
\psfrag{M}{$M^2_{\gamma \rho}$}
\psfrag{GPD}{$\!GPD$}
\centerline{
\raisebox{1.6cm}{\includegraphics[width=14pc]{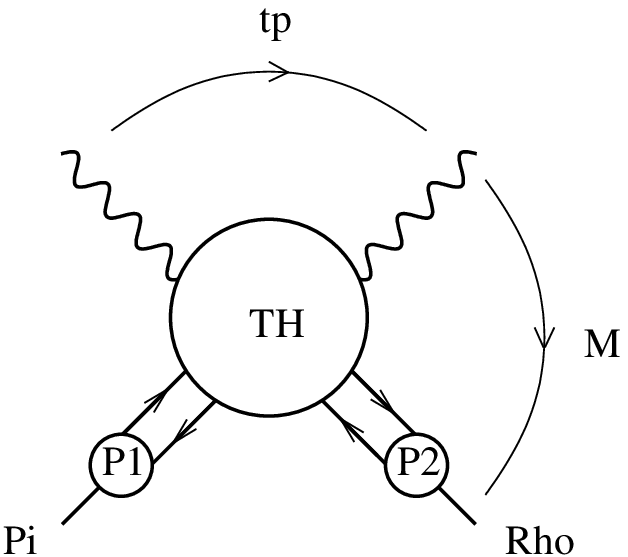}}~~~~~~~~~~~~~~
\psfrag{TH}{$\,\Large T_H$}
\includegraphics[width=14pc]{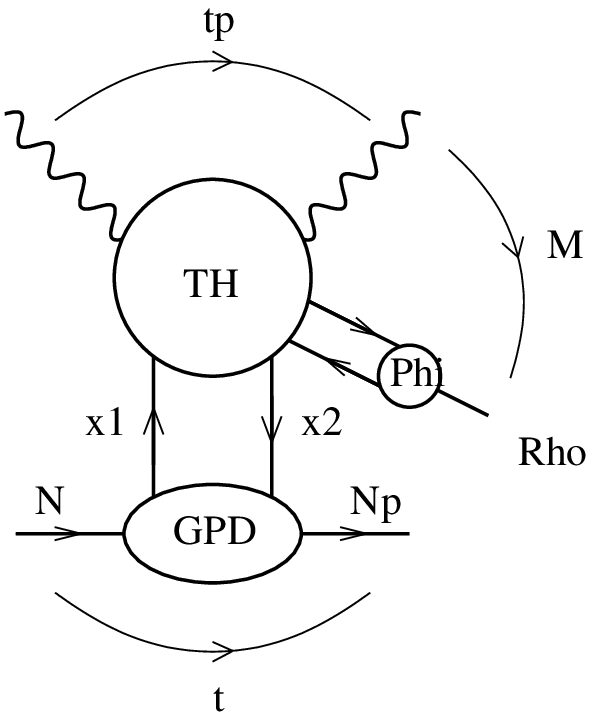}}
\caption{\label{process}The wide angle Compton scattering process (left) and its generalization to the photoproduction of a $\gamma \rho$ pair (right). }
\end{figure}

To factorize the amplitude of this process we use  the now classical proof of the factorization of exclusive scattering at fixed angle and large energy~\cite{LB}. The amplitude for the wide angle Compton scattering process $\gamma + \pi \rightarrow \gamma + \rho $ is written \cite{Nizic} as the convolution of mesonic DAs  and a hard scattering subprocess amplitude $\gamma +( q + \bar q) \rightarrow \gamma + (q + \bar q) $ with the final meson states replaced by a collinear quark-antiquark pair. 
We then extract from the factorization procedure of the deeply virtual Compton scattering amplitude near the forward region the right to replace one entering meson DA by a $N \to N'$ GPD, and thus get Fig.~1 (right panel). 
The needed skewness parameter $\xi$  is written in terms of the final photon - meson  squared invariant mass
$M^2_{\gamma\rho}$ as
\begin{equation}
\label{skewedness}
\xi = \frac{\tau}{2-\tau} ~~~~,~~~~\tau =
\frac{M^2_{\gamma\rho}-t}{S_{\gamma N}-M^2}\,.
\end{equation}

Indeed the same collinear factorization property bases the validity of the leading twist approximation which either replaces the meson wave function by its DA or the $N \to N'$ transition by nucleon GPDs. A slight difference is that light cone fractions ($z, 1- z$) leaving the DA are positive, while the corresponding fractions ($x+\xi,\xi-x$) may be positive or negative in the case of the GPD. Our  Born order calculation  shows that this difference does not ruin the factorization property.

In order for the leading twist factorization of a partonic amplitude to be valid, one should avoid the dangerous
kinematical regions where a small momentum transfer is exchanged in the
upper blob, namely small $t' =(k-q)^2$ or small $u'=(p_\rho-q)^2$, and the resonance regions for each  of the
invariant squared masses {$(p_\rho +p_{N'})^2,$} $(k+p_\rho)^2\,.$

Let us finally stress that our discussion applies as well to the case of electroproduction where a moderate virtuality of the initial photon may help to access the perturbative domain with a lower value of the hard scale $M_{\gamma \rho}$.

\section{Kinematics}

Our conventions  are the following. We decompose all momenta on a Sudakov basis  as $
v^\mu = a \, n^\mu + b \, p^\mu + v_\bot^\mu $,
with $p$ and $n$ the light-cone vectors
$
p^\mu = \frac{\sqrt{s}}{2}(1,0,0,1), n^\mu = \frac{\sqrt{s}}{2}(1,0,0,-1), $
$v_\bot^\mu = (0,v^x,v^y,0) $ and $v_\bot^2 = -\vec{v}_t^2\,.
$
The particle momenta read
\begin{equation}
\label{impini}
 p_1^\mu = (1+\xi)\,p^\mu + \frac{M^2}{s(1+\xi)}\,n^\mu~, \quad p_2^\mu = (1-\xi)\,p^\mu + \frac{M^2+\vec{\Delta}^2_t}{s(1-\xi)}n^\mu + \Delta^\mu_\bot\,, \quad q^\mu = n^\mu ~,
\end{equation}
\beqa
\label{impfinc}
k^\mu = \alpha \, n^\mu + \frac{(\vec{p}_t-\vec\Delta_t/2)^2}{\alpha s}\,p^\mu + p_\bot^\mu -\frac{\Delta^\mu_\bot}{2},~ ~p_\rho^\mu = \alpha_\rho \, n^\mu + \frac{(\vec{p}_t+\vec\Delta_t/2)^2+m^2_\rho}{\alpha_\rho s}\,p^\mu - p_\bot^\mu-\frac{\Delta^\mu_\bot}{2},\nonumber
\eqa
with $\bar{\alpha} = 1 - \alpha$ and $M$,  $m_\rho$ are the masses of the nucleon  and the $\rho$ meson.
The total center-of-mass energy squared of the $\gamma$-N system is
\begin{equation}
\label{energysquared}
S_{\gamma N} = (q + p_1)^2 = (1+\xi)s + M^2\,.
\end{equation}
From these kinematical relations it follows that :
\beq
\label{2xi}
2 \, \xi = \frac{(\pv -\frac{1}2 \dv)^2 }{s \, \alpha} +\frac{(\pv +\frac{1}2 \dv)^2 + \mr^2}{s \, \ar}\,,
\eq
and
\beq
\label{exp_alpha}
1-\alpha-\ar = \frac{2 \, \xi \, M^2}{s \, (1-\xi^2)} + \frac{\dv^2}{s \, (1-\xi)}\,.
\eq
On the nucleon side, the transferred squared momentum is
\begin{equation}
\label{transfmom}
t = (p_2 - p_1)^2 = -\frac{1+\xi}{1-\xi}\vec{\Delta}_t^2 -\frac{4\xi^2M^2}{1-\xi^2}\,.
\end{equation}
The other various Mandelstam invariants read
\begin{eqnarray}
\label{M_pi_rho}
s'&=& ~(p_\gamma +p_\rho)^2 = ~M_{\gamma\rho}^2= 2 \xi \, s \left(1 - \frac{ 2 \, \xi \, M^2}{s (1-\xi^2)}  \right) - \dv^2 \frac{1+\xi}{1-\xi}\,, \\
\label{t'}
- t'&=& -(p_\gamma -q)^2 =~\frac{(\vec p_t-\vec\Delta_t/2)^2}{\alpha} \;,\\
\label{u'}
- u'&=&- (p_\rho-q)^2= ~\frac{(\vec p_t+\vec\Delta_t/2)^2+(1-\alpha_\rho)\, m_\rho^2}{\alpha_\rho}
 \; .
\end{eqnarray}
Let us remind the reader that we are interested in the kinematical domain where $s', -t', -u'$ are large (as compared to $\Lambda^2_{QCD}$) and that $0 < \alpha, \alpha_\rho < 1$.

\section{The scattering amplitude}
\label{Sec:scattering}
The scattering amplitude of the process (\ref{process}) is written in the factorized form:
\begin{equation}
\label{ampl}
\mathcal{A}(t,M^2_{\gamma\rho},u')  = \sum\limits_{q,i} \int_{-1}^1dx\int_0^1dz\ T_i^q(x,v,z) \, H_i^{q}(x,\xi,t)\Phi_{\rho_{L,T}}(z)\,,
\end{equation}
where
$T_i^q$ is the hard part of the amplitude and $H_i^{q}$ the corresponding (chiral-even and chiral-odd)
GPDs of a parton $q$  in the nucleon target, and $\Phi_{\rho_{L,T}}(z)$ the leading twist chiral-even (resp. chiral-odd) distribution amplitude of the $\rho_L$ (resp. $\rho_T$) meson.

\begin{figure}[h]
\centerline{\includegraphics[width=32pc]{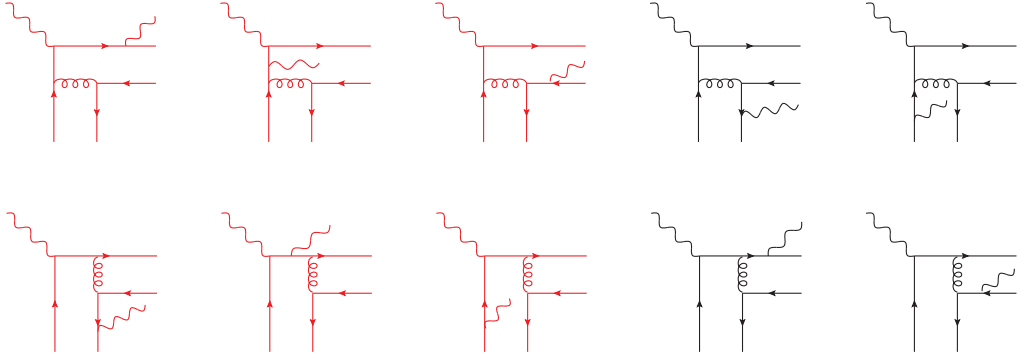}}
\caption{\label{diagrams}The Feynman diagrams describing the subprocess at leading order; in Feynman gauge only the 4 diagrams on the right contribute to the $\rho_T$ case}
\end{figure}
The scattering sub-process is described by
20 Feynman diagrams, but an interesting (quark-antiquark interchange) symmetry allows to deduce the contribution of half of the diagrams from the 10 diagrams shown on Fig.~\ref{diagrams} through a ($x \leftrightarrow -x \,; \,z \leftrightarrow 1-z$) interchange. Moreover, in Feynman gauge, only the 4 diagrams on the right of Fig.~\ref{diagrams} contribute to the chiral-odd case.

The scattering amplitudes get both  real and  imaginary parts. Focusing on the chiral-odd amplitude (since accessing transversity GPDs was the first motivation of our study), we get the following results. The $z$ and $x$ dependence  of this amplitude can be factorized as
\begin{equation}
 T_i^q = e_q^2 \,\alpha_{em}\, \alpha_s \,{\mathcal{N} (z,x)}\, \mathcal{T}^i
 \end{equation}
  with (in the gauge $p.\epsilon_k =0$):
\begin{eqnarray} 
\mathcal{T}^i &=& (1-\alpha) \left[ \left( \epsilon_{q\bot} . p_\bot \right) \left( \epsilon_{k\bot}.\epsilon_{\rho\bot} \right) - \left( \epsilon_{k\bot} . p_\bot \right) \left( \epsilon_{q\bot}.\epsilon_{\rho\bot} \right) \right] p_\bot^i \nonumber \\ 
&-& (1+\alpha) \left(\epsilon_{\rho\bot}.p_\bot\right) \left( \epsilon_{k\bot}.\epsilon_{q\bot}\right) p_\bot^i + \alpha \left( \alpha^2 -1\right) \xi s \left(\epsilon_{q\bot}.\epsilon_{k\bot}\right)\epsilon_\rho^i \\ 
&-&\alpha \left( \alpha^2 -1 \right) \xi s \left[ \left(\epsilon_{q\bot}.\epsilon_{\rho\bot}\right) \epsilon_{k\bot}^i - \left(\epsilon_{k\bot}.\epsilon_{\rho\bot}\right) \epsilon_{q\bot}^i \right]\,. \nonumber
\end{eqnarray}
Using as a first estimate the asymptotic form of the $\rho-$meson distribution amplitude, we perform analytically the integration over $z$. Inserting a model for the transversity GPDs \cite{PLB}, we use numerical methods for the integration over $x$.

Starting with the expression of the scattering amplitude (\ref{ampl}),
the differential cross-section as a function of $t$, $M^2_{\gamma\rho},$ $-u'$  reads
\begin{equation}
\label{difcrosec}
\left.\frac{d\sigma}{dt \,du' \, dM^2_{\gamma\rho}}\right|_{\ t=t_{min}} = \frac{|\mathcal{M}|^2}{32S_{\gamma N}^2M^2_{\gamma\rho}(2\pi)^3}.
\end{equation}
\noindent
We show in Fig.~\ref{resultS20}  this cross section (\ref{difcrosec}) as a function of  $-u'$ at $S_{\gamma N}$ = 20 GeV$^2$  for  $M^2_{\gamma\rho}$ = 6 GeV$^2$ i.e. $\xi = 0.186$, with cuts in $-u'$ corresponding to the constraints $-t'>1$ GeV$^2$ and $-u'> 1$ GeV$^2$. The cross section grows with $(-u')$ but its normalization is rather small. We expect a larger cross-section for the longitudinal $\rho$ case where chiral-even GPDs contribute; this will not help disentangling the transverse $\rho$ cross section, although a complete analysis of the angular distribution of the emerging $\pi^+ \pi^-$ pair allows in principle to access the chiral-odd sensitive contribution at the amplitude level.
\begin{figure}[h]
\vspace{.5cm}
\psfrag{U}{\raisebox{-.15cm}{$1$}}
\psfrag{D}{\raisebox{-.15cm}{$2$}}
\psfrag{T}{\raisebox{-.15cm}{$3$}}
\psfrag{Q}{\raisebox{-.15cm}{$4$}}
\psfrag{C}{\raisebox{-.15cm}{$5$}}
\psfrag{V}{\raisebox{0cm}{\hspace{-.4cm}$0.1$}}
\psfrag{W}{\raisebox{0cm}{\hspace{-.4cm}$0.2$}}
\psfrag{B}{\raisebox{.3cm}{\hspace{-.9cm}$\left.\frac{d \sigma}{dt du' dM_{\gamma \rho}^2}\right|_{|t|_{\rm min}}$(pb.GeV$^{-6}$)}}
\psfrag{A}{\raisebox{-.3cm}{$-u'$ (GeV$^2$)}}
\centerline{\hspace{-.5cm}\includegraphics[width=12cm]{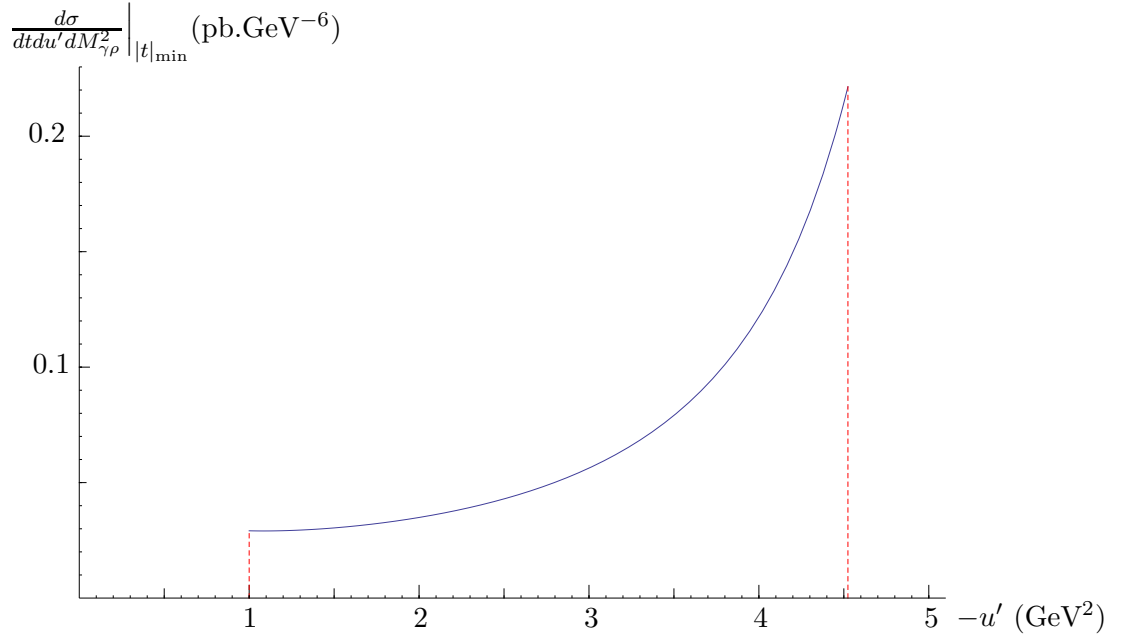}}
\caption{\label{resultS20}The differential cross section (\ref{difcrosec}) for the production of $\gamma \rho_T$ involving chiral-odd GPDs, as a function of  $-u'$ at $S_{\gamma N}$ = 20 GeV$^2$  for $M^2_{\gamma\rho}$ =  6 GeV$^2$, i.e. $\xi = 0.186$.}
\label{resultS20}
\end{figure}

The quest for an easy extraction of chiral-odd GPDs is obviously not solved by  our proposal for $\gamma \rho_T$ photoproduction.

 \ack
 This work is partly supported by the Polish Grant NCN No DEC-2011/01/B/ST2/03915 and  the French grant ANR PARTONS (Grant No.ANR-12-MONU-0008-01). L.~S. was partially supported by a French Government Scholarship.

\end{document}